%% file: AnonymousSubmission2027.tex
\documentclass[letterpaper]{article} 
\usepackage{aaai2027}  
\usepackage[hyphens]{url}  
\usepackage{graphicx} 
\usepackage{natbib}  
\usepackage{caption} 
\usepackage{algorithm}
\usepackage{algorithmic}

\input{package}

\usepackage{xspace}
\usepackage{newfloat}
\usepackage{listings}
\DeclareCaptionStyle{ruled}{labelfont=normalfont,labelsep=colon,strut=off} 
\floatstyle{ruled}
\newfloat{listing}{tb}{lst}{}
\floatname{listing}{Listing}

\usepackage{booktabs}

\title{TransMem: Transforming Hidden States into Memory for Large Language Models}
\author{
    Haodong Lei\textsuperscript{\rm 1,2},
    Junming Liu\textsuperscript{\rm 2},
    Yirong Chen\textsuperscript{\rm 2},
    Pinlong Cai\textsuperscript{\rm 2},
    Botian Shi\textsuperscript{\rm 2},
    Ding Wang\textsuperscript{\rm 2}\corresponding,
    Hongsong Wang\textsuperscript{\rm 1}\corresponding
}

\affiliations{
    \textsuperscript{\rm 1}Southeast University, Nanjing, Jiangsu, China\\
    \textsuperscript{\rm 2}Shanghai Artificial Intelligence Laboratory, Shanghai, China\\
    leihaodong@seu.edu.cn,
    hongsongwang@seu.edu.cn
}

\begin{document}

\maketitle

\begin{abstract}
Large language model (LLM) agents increasingly operate over long interaction histories, where effective reasoning requires identifying and exploiting task-relevant evidence distributed across past observations and actions. However, useful information encoded in previously computed representations is often underutilized during subsequent generation. We propose \textbf{TransMem}, a lightweight inference-time parametric memory module that transforms sparse historical hidden states from a frozen LLM backbone into reusable memory representations. TransMem uses a lightweight gating network to dynamically apply the latent intervention to the current hidden states, without repeatedly encoding the preceding context. To learn transferable memory utilization rather than task-specific knowledge, we introduce evidence-conditioned self-distillation. A memory-augmented student processes the full context and matches the predictive distribution of an evidence-only teacher that shares the same frozen backbone. Experiments on LoCoMo, HotpotQA, and MemoryAgentBench demonstrate consistent improvements across different model architectures and scales. TransMem yields gains of 11.58--29.25 $F_1$ on LoCoMo and 10.20--13.03 $F_1$ on HotpotQA, while improving the average MemoryAgentBench accuracy from 29.54\% to 40.00\%. These results establish sparse historical hidden states as an effective and efficient memory substrate for long-context LLM agents. Our code is available at \url{https://github.com/Haodong-Lei-Ray/TransMem}.
\end{abstract}


\input{main}

\appendix

\clearpage
\section{Appendix}
\input{supp}
\clearpage

\bibliography{aaai2027}


\end{document}

%% file: package.tex
\usepackage{booktabs}
\usepackage{multirow}
\usepackage{graphicx}
\usepackage{subcaption}
\usepackage{listings}
\usepackage{amssymb}
\usepackage{amsmath}
\usepackage{tabularx}
\usepackage{xcolor}
\usepackage[table]{xcolor}

%% file: main.tex
\section{Introduction}
Large language models (LLMs) are increasingly expected to support sustained interactions over long contexts, particularly in applications such as coding agents~\cite{SWE-agent}, long-horizon agent systems~\cite{LBT, MemAgent}, and complex reasoning~\cite{agashe2025agent}.
The crucial information needed to answer a query may be distributed across a long interaction history or document~\cite{longmemeval}. The challenge is therefore not merely to accept longer inputs, but to identify and exploit the evidence that matters for the current prediction~\cite{du2025rethinkingmemoryllmbased}.

A straightforward response is to enlarge the context window or scale the backbone model~\cite{pmlr-v235-ding24i, zhao-etal-2025-helene}. However, model predictions can be sensitive to the position of relevant evidence, and long-context reasoning often degrades when useful information is distant from the generation point~\cite{liu-etal-2024-lost, hsieh-etal-2024-found}. 
External memory addresses this limitation by summarizing, retrieving, or storing information outside the current computation~\cite{memcot,yang2026cfgm}. Parametric memory encodes information into auxiliary latent spaces, which helps mitigate limitations imposed by the native computational structure of Transformers~\cite{wei2026mlpmemory,deltamem}. These mechanisms are useful when information must persist across tasks or sessions. However, they do not directly address a distinct test-time problem, which is how an LLM can better exploit the representations already computed for the current long context.

\begin{figure*}[t]
    \centering
    \includegraphics[width=1\textwidth]{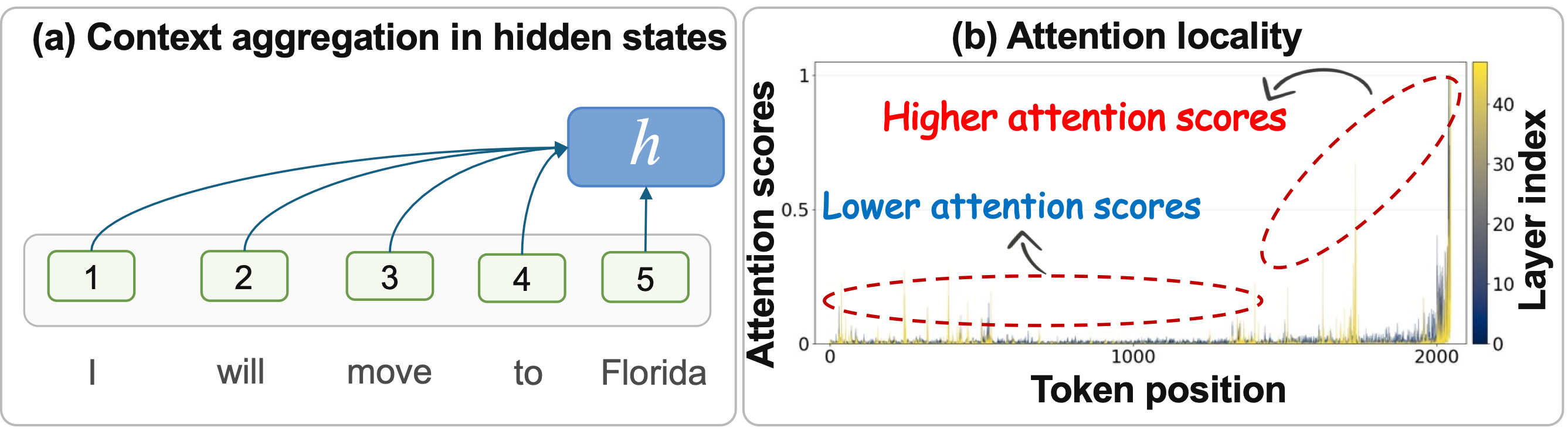}
    \caption{\textbf{Two properties of contextual representations in LLMs.}
(a) \textit{Hidden states aggregate preceding context}. Owing to the autoregressive computation of causal Transformers, each hidden state is computed from its preceding tokens. Consequently, later hidden states can serve as contextual representations of the preceding sequence.
(b) \textit{Transformer representations are region-sensitive.} Excluding the initial sink position, evidence appearing later in the context generally receives higher attention, reflecting a recency bias.}
    \label{fig:intro}
\end{figure*}

Based on this question, we observe two properties of causal transformers. \textbf{(1)} \textbf{Hidden states naturally aggregate information} from their preceding tokens~\cite{li2024eagle,deepseekai2025deepseekv3technicalreport}. As shown in Figure~\ref{fig:intro}(a), a hidden state near the end of a context segment can therefore serve as a compact representation of that segment and its prefix. \textbf{(2)} \textbf{These representations are position-sensitive.} Because attention tends to favor nearby tokens~\cite{liu-etal-2024-lost}, representations formed at different positions preserve different parts of the context with different fidelity.
This effect becomes more pronounced as the context length increases. As illustrated in Figure~\ref{fig:intro}(b), even when the information most relevant to the query appears earlier in the context, the Transformer may still assign higher attention scores to more recent tokens. Consequently, information from a long context is distributed across hidden states at different positions, rather than being fully retained by representations near the current generation position.
These properties suggest that a small set of strategically selected hidden states can provide complementary views of a long input, without recomputing the entire long context as another memory. Consequently, substantial task-relevant information may remain latent in previously computed hidden states but underutilized during subsequent generation.

Based on these observations, we propose TransMem, an inference-time parametric memory module that transforms sparse historical hidden states into task-specific memory representations to enhance long-context reasoning. Rather than learning to store particular knowledge representations, TransMem learns the memory capability itself through an evidence-conditioned self-distillation training method. Specifically, it learns how to extract useful information from existing hidden states to answer the current query. As a result, TransMem does not memorize knowledge tied to a specific long context, but instead acquires a general capability for processing and reasoning over long contexts. It offers a scalable route toward improving memory capacity independently of the backbone model size.


Our contributions are summarized as follows:
\begin{itemize}
    \item We propose \textbf{TransMem, a lightweight inference-time parametric memory module} that transforms previously computed sparse hidden states from a frozen LLM backbone into reusable memory representations.
    \item We introduce \textbf{an evidence-conditioned self-distillation} training method. By exposing the teacher and student to different input contexts, the parameterized memory module learns to recover evidence-conditioned predictions from full long contexts. 
    \item Extensive experiments on LoCoMo~\cite{LOCOMO}, HotpotQA~\cite{yang2018hotpotqa}, and MemoryAgentBench~\cite{hu2026evaluating} have shown consistent improvement on long-context reasoning across backbone architectures and scales.
\end{itemize}

\section{Related Work}
\label{sec:related_work}
\subsection{External Memory for LLMs}
The integration of long-term memory mechanisms~\cite{SurveyMemory, srivastava2026effgen} marks a fundamental paradigm shift in the evolution of LLM agents~\cite{SurveyMemory}. Early frameworks like MemGPT~\cite{memgpt} pioneered operating-system-inspired paging and segmentation to manage extended contexts. Building upon this, scalable architectures like Mem0~\cite{mem0} dynamically consolidate memory states to mitigate the severe limitations of fixed context windows. Retrieval-augmented memory systems frequently suffer from critical information omissions and semantic noise~\cite{latimer2025hindsight2020buildingagent}. To address this issue, recent studies have explored iterative retrieval paradigms, including MemCoT~\cite{memcot}, MemEvolve~\cite{MemEvolve}, SimpleMem~\cite{SimpleMem}, and ReasoningBank~\cite{ouyang2026reasoningbank}. These approaches substantially improve retrieval accuracy by enabling progressive evidence refinement and adaptive memory exploration. However, they often suffer from limited robustness and increased computational overhead due to the repeated retrieval process~\cite{srivastava-etal-2025-thinkslm}. Moreover, these methods fundamentally do not eliminate the information loss introduced during memory extraction. Instead, they remain vulnerable to premature termination within the retrieval loop, which can prevent the model from locating the necessary evidence and ultimately lead to retrieval failure.

\subsection{Parametric Memory for LLMs}
Compared with memory mechanisms that store information in external knowledge repositories, parameterized memory provides a more expressive representation space by directly encoding knowledge into model parameters. Therefore, it represents a more advanced paradigm for enhancing the memory capacity of LLMs. Early attempts along this direction typically employ auxiliary models to explicitly or implicitly encode additional information, including MeMo~\cite{MeMo}, MemVerse~\cite{memverse}, Memory Decoder~\cite{cao2025memorydecoder}, MemSifter~\cite{MemSifter}, MEMTS~\cite{MEMTS}, and Mem-$\pi$~\cite{wang2026mempiadaptivememorylearning}. More sophisticated approaches instead design specialized architectures that introduce dedicated representations to enhance long-context reasoning capabilities~\cite{memgen, ji2026parametricmemorydecodingzeroshot}. For example, DRIFT~\cite{DRIFT} reveals that the last-layer feature may already contain highly compressed information from long contexts, suggesting that intermediate representations can serve as effective memory carriers. Meanwhile, the decoupling of memory capability from reasoning capability has emerged as an increasingly important research direction. MLPMemory~\cite{wei2026mlpmemory} maps last-layer features into a latent distribution to inject long-context reasoning ability, while $\delta$-mem~\cite{deltamem} adopts a heterogeneous delta-network architecture to incorporate long-context reasoning capabilities into layer-wise representations.
\begin{figure*}[t]
    \centering
    \includegraphics[width=1\textwidth]{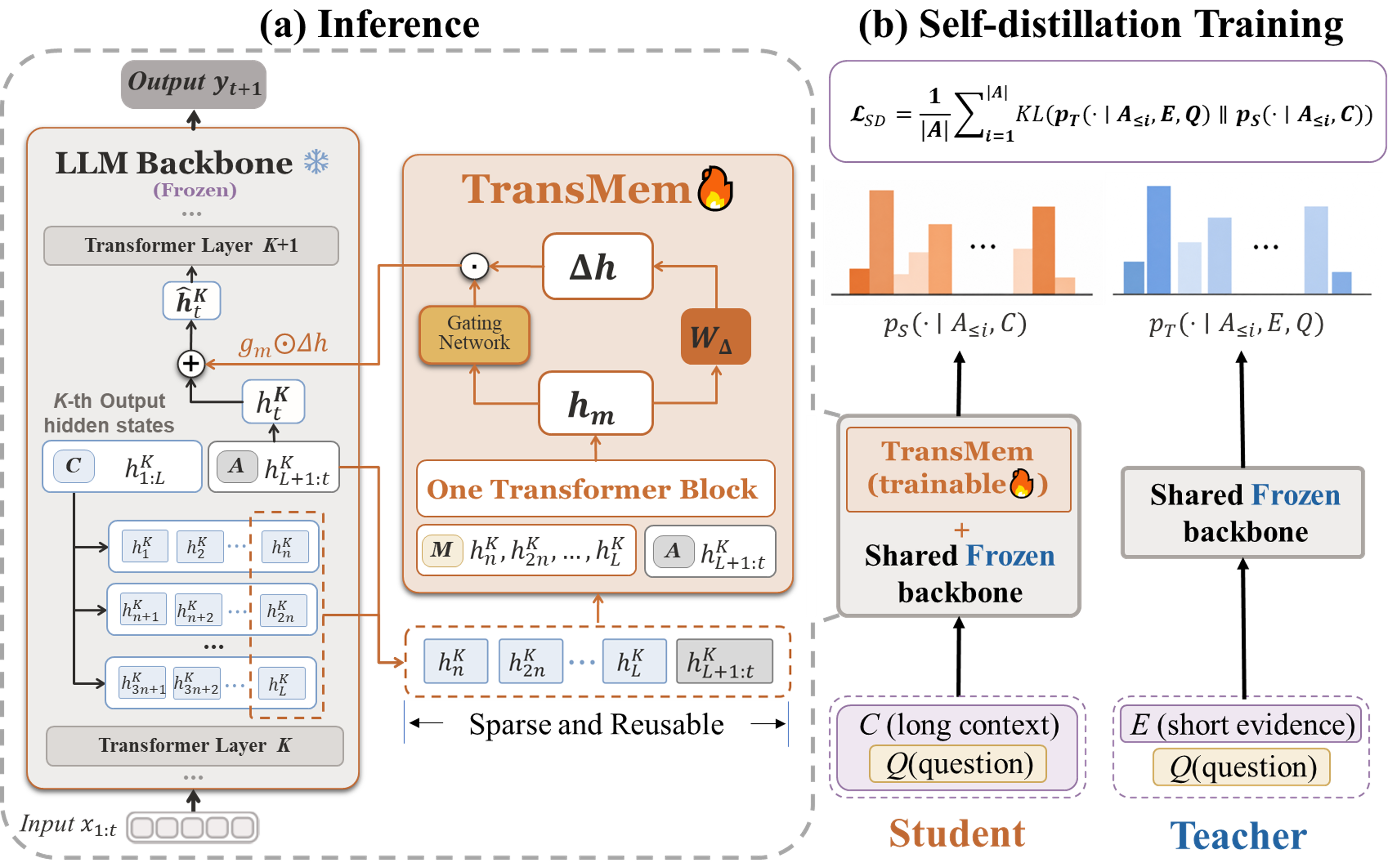}
    \caption{The overall architecture of the TransMem framework.
    (a) Inference with TransMem. TransMem is applied after the frozen LLM backbone produces the hidden states at the $K$-th layer. The selected historical hidden states are transformed into memory representations by the TransMem module.
    (b) Evidence-conditioned self-distillation. The student model consists of the frozen LLM backbone augmented with TransMem and receives the full long context $C$. The teacher model shares the same frozen backbone parameters but receives only the evidence $E$ and question $Q$, providing an evidence-conditioned supervision signal. The student is trained to recover the teacher prediction distribution from the full long context.
    }
    \label{fig:main}
\end{figure*}

\section{Method}
\label{sec:method}
In this section, we present the TransMem framework, including its inference and training procedures.
Figure~\ref{fig:main} provides an overview of the framework. During inference, the frozen backbone performs its standard forward computation, while TransMem abstracts a small set of historical hidden states into a memory shift. This shift influences current output hidden states to better preserve key information from long contexts. We train TransMem through self-distillation to recover the predictive distribution of a teacher model conditioned on gold evidence, enabling the memory module to preserve crucial information while suppressing contextual noise.

\subsection{Preliminaries}
During the prefill stage, the input long context is denoted as C. Let $Q\subseteq C$ denote the question and $E\subseteq C$ denote the evidence that supports the answer. At position $t$, the tokenized input is $x_{1:t}=[C;A_{ \leq t}]$, where $C=x_{1:L}$ is the long context and $A_{\leq t}=(y_{L+1},\ldots,y_{t})$ is the generated answer prefix. The next token is $y_{t+1}$. The hidden states produced by the $K$-th layer for $x_{1:t}$ are: 
\begin{equation}
h^K_{1:t}=[H_C;H_A]=[h^K_{1:L};h^K_{L+1:t}],
\end{equation}
where $h_{t}^K \in \mathbb{R}^{1\times d}$, with $d$ denoting the hidden dimension of the frozen backbone model. The context features $H_C$ are computed during prefill, during which TransMem does not intervene in generation.

With the KV cache, each layer only needs to receive the final hidden state at
each decoding step. In a standard LLM forward pass, transformer layer $K+1$
produces the hidden state of the next layer as
\begin{equation}
  h^{K+1}_{t}
  =f^{K+1}_{\theta}(h^K_{t}),
  \label{eq:ori}
\end{equation}
where $f^{K+1}_{\theta}$ represents the $(K+1)$-th Transformer block.

\subsection{Evidence-Conditioned Self-distillation Training}
\label{sub:inloop}
We train TransMem with self-distillation. Our goal is to improve the ability of LLMs to identify and exploit query-relevant evidence from long contexts. Therefore, instead of training the model to memorize the entire context, we aim to encourage the model to suppress irrelevant information and focus on evidence that supports the answer. To achieve this goal, we construct the teacher by providing only the evidence $E$ and query $Q$. Since the teacher has access to the evidence without irrelevant context noise, its output distribution represents a great behavior on evidence. In contrast, the student is equipped with learnable TransMem. It receives the full context $C$ and learns to recover the teacher's evidence-conditioned predictions through distillation.

The teacher runs the frozen backbone on $(E,Q)$, while the student runs the TransMem-augmented backbone on $C$. The teacher generates the answer trajectory $A$, and the student follows its answer prefix during teacher forcing. Both paths share the same frozen LLM backbone, so no separate teacher model is introduced. Let $p_T(\cdot\mid A_{<i}, E, Q)$ denote the teacher distribution along its answer trajectory. A student model augments the frozen backbone with TransMem and produces $p_S(\cdot\mid A_{<i}, C)$.
As shown in Figure~\ref{fig:main}(b), the training objective is the forward KL divergence at the language model output. We optimize the TransMem module by minimizing the forward KL divergence between the teacher and student distributions:
\begin{equation}
  \mathcal{L}_{\mathrm{SD}}
  = \frac{1}{|A|}\sum_{i=1}^{|A|}
    \mathrm{KL}\!\left(
      p_T(\cdot\mid A_{<i}, E, Q)\,\middle\|\,p_S(\cdot\mid A_{<i}, C)
    \right).
  \label{eq:tfkl}
\end{equation}

Since the teacher provides soft supervision, we use the teacher distribution as the target distribution. The student distribution is read through the frozen language model head. Gradients from Eq.~\eqref{eq:tfkl} pass through the frozen upper layers into every lower TransMem module. The backbone receives no parameter updates, while each correction is optimized for its final token-level effect.

A correction at layer $K$ changes the input distribution of every higher
layer. Training modules against independently cached layer targets therefore
creates a mismatch once lower-layer corrections become active. In-loop training keeps the frozen backbone in the computation graph and optimizes all memory modules jointly. The resulting upper-layer states include the corrections from every active lower module~\cite{zhang2026opsdlonpolicyselfdistillationlongcontext}. 

\subsection{Inference with TransMem}
\label{sub:setup}
The TransMem module at the $K$-th layer consists of a Transformer block $f_{m}^{K}$, a learnable projection $W^{K}_{\Delta}$, and a learnable gating network $f_{g}^{K}$.

TransMem applies its correction only while generating $A_{\leq t}$. Once
$h^K_{t}$ is obtained, we divide $H_C$ into segments of size $n$ and
use the final hidden state of each segment as a memory feature. This gives
$H_M={h^K_{n},h^K_{2n},...,h^K_{L}}$. The size of $H_M$ remains small, with the optimal value being only 5, which ensures the sparsity of the memory representation.

As illustrated in Figure~\ref{fig:main}(a), we feed $H_M$ and $H_A$ into
TransMem. They first undergo standard autoregressive attention to produce the
memory feature $h_m$ from the different segments. A learnable projection
$W^{K}_{\Delta} \in \mathbb{R}^{d \times d}$ then transforms $h_m$ into the final memory shift $\Delta h_m$.
The complete process is
\begin{equation}
h_m=f^{K}_m(H_M,h^A_{1:t}),
\end{equation}
\begin{equation}
\Delta h =W^{K}_{\Delta} h_m.
  \label{eq:transmem}
\end{equation}
Not every memory shift $\Delta h$ is useful, and noisy shifts may adversely affect the output. Therefore, we use the learnable gating network $f^{K}_g$ to scale the memory shift.
Specifically,
\begin{equation}
g_m =f^{K}_g(h_m)=
\alpha \cdot
\sigma
\left(
\frac{W_g h_m}{\tau}
\right),
  \label{eq:gate}
\end{equation}
where $\sigma(x)=\frac{1}{1+e^{-x}}$, $\alpha=2$, $\tau$ is gate temperature, and $W_g\in \mathbb{R}^{d \times d}$ is learnable gate weight. The core correction formula is
\begin{equation}
  \hat h^K_{t}  = h^K_{t}+g_m\odot\Delta h,
  \label{eq:inject}
\end{equation}
where $t \geq L+1$, indicating that the memory correction is
applied only during answer generation.
Inference with the TransMem memory correction then becomes:
\begin{equation}
  h^{K+1}_{t}
  =f^{K+1}_{\theta}(\hat h^K_{t}).
  \label{eq:inject}
\end{equation}
Therefore, our method enhances long-context reasoning through a lightweight network without extensively recomputing long text segments. Our method does not store additional information from the preceding context. Instead, its contextual awareness relies on the already computed vectors in $H_M$. The features in $H_M$ can serve as memory because of the computational properties of a transformer. In particular, $h^K_L$ encodes transformed information from $x_{1:L}$ and therefore acts as a feature representation of that prefix. A similar operation can also be found in EAGLE~\cite{li2024eagle} and DeepSeek MTP~\cite{deepseekai2025deepseekv3technicalreport}.

In this sense, the method parameterizes part of the transformer's memory
capability in an independent network. It therefore takes a further step toward
separating the reasoning and memory capabilities of an LLM.

TransMem is not applied to all layers. Empirically, we find that inserting memory modules into only the last four layers of the LLM is sufficient to achieve strong performance. It further makes our method more lightweight. Furthermore, unlike existing latent memory methods, TransMem is invoked only when the model begins generating the answer sequence $A$. This sparse inference strategy further reduces the additional computation.

To construct sparse memory representations, we adopt a dynamic segmentation strategy. Specifically, we partition the long context into $B$ segments and set the segment length to $n = \frac{L}{B}$. We note that this choice is not necessarily optimal, and the segment length can be adapted to different tasks or application scenarios.

When the long context $C$ exceeds the maximum context window of the frozen backbone model, we adopt a chunk-wise memory extraction strategy. Specifically, $C$ is divided into $S$ chunks according to the backbone context window, denoted as $\{C_1,\ldots,C_S\}$. Each chunk independently produces a sparse memory representation $H_M^i$. During answer generation, the final memory is formed by aggregating all chunk-level memories:
\[
H_M = \operatorname{Concat}(H_M^1, \ldots, H_M^S).
\]
This strategy enables TransMem to exploit information beyond the backbone context window while introducing only a small amount of additional computation.

\section{Experiments}
\label{sec:experiments}
\input{tab/main}
\input{tab/MAB}

\subsection{Experimental Setup}
\label{sub:exp_setup}
\paragraph{Evaluation and Benchmarks. }
We evaluate both general reasoning ability and memory effectiveness. We evaluate multi-hop question answering performance on the HotpotQA test set~\cite{yang2018hotpotqa}, which assesses multi-hop reasoning. Memory effectiveness is evaluated on LoCoMo \cite{LOCOMO} and MemoryAgentBench \cite{hu2026evaluating}, which measure the retention, retrieval, and use of information over extended interaction histories.
LoCoMo contains contexts with an average length of approximately 16K tokens, making it suitable for evaluating long-context reasoning capabilities. In contrast, MemoryAgentBench includes many scenarios with context lengths exceeding 256K tokens, providing a more challenging evaluation of latent-space memory utilization. Following mem0 \cite{mem0}, we exclude the adversarial question category from LoCoMo. LoCoMo comprises 841 single-hop (SH), 282 multi-hop (MH), 96 open-domain (OD), and 321 temporal (TP) reasoning challenges.
\\

\paragraph{Baselines. }
We compare TransMem with representative memory approaches based on different memory mechanisms, including external memory and parametric memory methods. For external memory methods, we include BM25 RAG \cite{lewis2020retrieval}, which retrieves relevant passages from historical contexts and appends them to the input, LLMLingua-2 \cite{pan-etal-2024-llmlingua}, which compresses long contexts into shorter textual representations, MemoryBank \cite{MemoryBank}, Memory-R1-GRPO \cite{yan-etal-2026-memory} and TriMem \cite{sun2026trimem}, which maintain historical information as retrievable textual memories. For parametric memory methods, we compare with Context2LoRA \cite{back2026understandingloraknowledgememory}, MemGen \cite{memgen}, MLP Memory \cite{wei2026mlpmemory} and $\delta$-Mem~\cite{deltamem},  which introduce additional trainable components to encode or transform contextual information into memory representations. All experiments are conducted with a fixed random seed of 42.
\\

\paragraph{Implementation Details. }
To evaluate the generality of TransMem, we conduct experiments on LLM backbones with different architectures and parameter scales, including Qwen3-4B-Instruct~\cite{qwen3technicalreport}, Qwen2.5-14B-Instruct~\cite{qwen2025qwen25technicalreport}, and Llama3.1-8B-Instruct~\cite{grattafiori2024llama3herdmodels}. TransMem is a learnable memory component, which is trained only on the HotpotQA training set. We follow the same data processing strategy as AgeMem~\cite{AgeMem} to ensure the effective utilization of the HotpotQA training set. Notably, we ensure that the HotpotQA training set contains no content overlapping with the test set, thereby avoiding unfair performance gains due to test-set leakage. TransMem is inserted into the final four transformer layers of the student LLM model. Its transformer block consists of a single layer that follows the Qwen3 transformer architecture. Training uses one NVIDIA A100 Tensor Core GPU and 15 gradient accumulation steps per rank, resulting in a global batch size of 30.

\subsection{Main Results}
We evaluate TransMem on models with different scales and architectures. As shown in Table~\ref{tab:main}, TransMem consistently improves long-context reasoning performance across all evaluated models. On LoCoMo, it achieves relative improvements of at least 27\% over the corresponding backbone models. On HotpotQA, TransMem yields relative EM improvements of approximately 14\%. We further observe that the effectiveness of TransMem may depend on the underlying capability of the backbone model. For example, although Qwen3-4B-Instruct has fewer parameters than Qwen2.5-14B-Instruct, its more advanced architecture enables comparable performance on LoCoMo. Moreover, TransMem demonstrates robustness to substantial changes in the backbone architecture. Although Llama3.1-8B-Instruct does not achieve performance comparable to the other two backbones, the transformer block within TransMem adopts the more advanced Qwen3 architecture, allowing it to achieve an $F_1$ score of 51.64 on LoCoMo and 71.63 on HotpotQA. These results are comparable to those of Qwen2.5-14B-Instruct with a larger parameter scale.

As shown in Table~\ref{tab:MAB}, we further evaluate TransMem on Memory Agent Bench, which contains substantially longer contexts than LoCoMo, with many examples exceeding 256K tokens. Since TransMem reuses historical hidden representations that have been truncated from the current context window, it can preserve information beyond the effective context length of the backbone model. As a result, it consistently improves performance across different memory settings, demonstrating that the proposed memory mechanism remains effective under ultra-long-context reasoning.

Notably, although TransMem is trained only on the HotpotQA training set, it generalizes effectively to both LoCoMo and Memory Agent Bench. This phenomenon demonstrates that the learned memory capability transfers well across diverse long-context reasoning tasks, rather than memorizing knowledge specific to the HotpotQA dataset.

\subsection{Ablation Studies}
\begin{table}[htbp]
    \centering
    \caption{Module Ablation Results on LoCoMo with Qwen3-4B-Instruct as the backbone.}
    \resizebox{0.48\textwidth}{!}{
    \begin{tabular}{lccccc}
    \toprule
    \textbf{Model} & 
\textbf{Avg.} &
\textbf{MH} &
\textbf{TP} &
\textbf{OD} &
\textbf{SH} \\
    \midrule
    TransMem & \textbf{54.11} & \textbf{52.04} & \textbf{45.62} & 20.64 & \textbf{61.87} \\
    w/o Gating Network & 53.09& 48.04& 45.09& \textbf{21.57}& 61.45\\
    w/o Transformer Block & 40.79 & 38.39 & 32.89 & 10.77 & 48.05 \\
    \bottomrule
    \end{tabular}
    }
    \label{tab:ablation}
\end{table}
To investigate the contribution of each component in TransMem, we conduct an ablation study by removing individual modules. As shown in Table~\ref{tab:ablation}, the results demonstrate that the proposed transformer block is the primary contributor to the overall performance of TransMem. Although the Gating Network provides a relatively modest improvement in the overall score, it yields a noticeably larger gain on multi-hop reasoning. This observation suggests that the gating mechanism facilitates the retrieval and utilization of relevant memory representations, thereby improving the model's ability to exploit stored information during complex reasoning.
\subsection{Further Study}
\begin{figure}[t]
    \centering
    \includegraphics[width=0.45\textwidth]{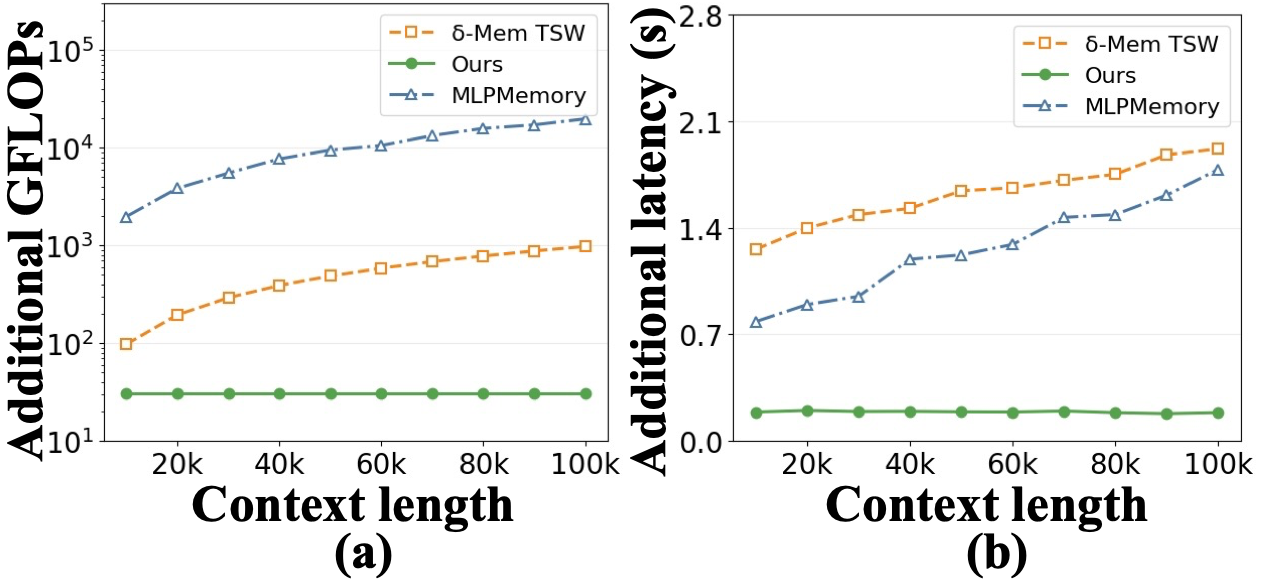}
    \caption{Efficiency comparison in terms of computation cost and inference latency. (a) Additional computational cost introduced by the memory module, shown on a logarithmic scale. (b) Average additional inference latency introduced by the memory module.}
    \label{fig:efficiency}
\end{figure}

\paragraph{Efficiency. }
To evaluate the efficiency of TransMem, we measure the additional computational cost and inference latency under different context lengths and latent memory configurations. We report the additional latency introduced by the memory module, excluding the frozen backbone prefill computation. As shown in Figure~\ref{fig:efficiency}(a), we compare the additional inference costs of TransMem, $\delta$-Mem, and MLPMemory on Qwen3-4B-Instruct as the context length increases from 10K to 100K tokens. $\delta$-mem introduces an approximately linear increase in computational overhead, where every additional 10k tokens in the latent memory leads to nearly 100 GFLOPs of extra computation. Although this overhead remains relatively small compared with the backbone model, the computational cost of TransMem is independent of the context length and only depends on the number of selected memory slots $B$, equivalently $|H_M|$. Therefore, TransMem maintains a constant computational overhead regardless of the input context length. Moreover, since TransMem is inserted into only a subset of backbone layers rather than all layers, its computational cost remains lower than $\delta$-mem even with a 10k-token context.

TransMem also achieves lower inference latency, as shown in Figure~\ref{fig:efficiency}(b). Unlike methods that repeatedly encode historical memory representations, TransMem performs sparse memory computation only during answer generation without additional processing of previous contexts. Therefore, increasing the context length does not introduce additional inference latency, demonstrating the scalability of TransMem for long-context reasoning.

\begin{table}[htbp]
    \centering
    \vspace{-5pt}
    \caption{\textbf{LoCoMo $F_1$ under different TransMem injection depths and
    layer positions.} Experiments are conducted on Qwen3-4B-Instruct.}
    \vspace{-5pt}
    \label{tab:injection_depth_position}
    \begin{tabular}{lccccc}
    \toprule
    \textbf{Model} & 
\textbf{Avg.} &
\textbf{MH} &
\textbf{TP} &
\textbf{OD} &
\textbf{SH} \\
    \midrule
    Last-2 & 53.14 & 49.91 & 44.70 & 18.35 & 61.41 \\
    Last-4 & \textbf{54.11} & \textbf{52.04} & \textbf{45.62}& \textbf{20.64} & \textbf{61.87} \\
    Last-8  & 53.27 & 50.29 & 44.31 & 20.44 & 61.43 \\
    Middle 28-31 & 52.85 & 50.18 & 43.74 & 20.07 & 60.96 \\
    Middle 22-25 & 52.32 & 47.24 & 43.26 & 20.67 & 61.09 \\
    Middle 18-21 & 51.54& 45.18 & 44.48 & 20.17 & 59.96 \\
    Middle 12-16 & 51.11& 44.00 & 42.92 & 20.57 & 60.11 \\
    \bottomrule
    \end{tabular}
    \label{tab:Depth}
\end{table}

\paragraph{Insertion Depth. }
We further investigate the impact of inserting TransMem at different layers. As illustrated in Table~\ref{tab:Depth}, we find that increasing the number of inserted layers does not always lead to better performance, and earlier inserted layers are not necessarily beneficial. The Last-2 to Last-8 settings denote inserting TransMem into the last 2 to 8 layers of the backbone, respectively. Since Qwen3-4B-Instruct contains 36 layers, the Middle settings insert TransMem into selected intermediate layers. We observe that inserting TransMem into the last four layers achieves the best performance, particularly on the Multi-Hop task. Compared with inserting TransMem into only the last two layers, Last-4 provides more effective memory enhancement. However, further increasing the number of inserted layers to Last-8 does not bring additional improvements. Similarly, inserting TransMem into layers 28 to 31 does not provide noticeable gains. Overall, inserting TransMem into the last layers generally yields better performance than inserting it into the middle layers. Earlier insertion locations may tend to provide weaker improvements, suggesting that effective memory formation may occur primarily in deeper layers.

\begin{table}[t]
    \centering
    \caption{Effect of different training objectives on LoCoMo $F_1$.
    Experiments are conducted on Qwen3-4B-Instruct. All training sets are from HotpotQA.}
    \vspace{-5pt}
    \label{tab:train}
    \begin{tabular}{lccccc}
    \toprule
    \textbf{Method} &
    \textbf{Avg.} &
    \textbf{MH} &
    \textbf{TP} &
    \textbf{OD} &
    \textbf{SH} \\
    \midrule
    SFT & 48.82 & 47.06 & 40.66 & 19.45 & 55.89 \\
    GRPO & 53.18 & 51.36 & 43.92 & 17.42 & 61.40 \\
    OPD & 52.80 & 51.17 & 45.01 & 18.96 & 60.19 \\
    ECSD & \textbf{54.11} & \textbf{52.04} & \textbf{45.62}
        & \textbf{20.64} & \textbf{61.87} \\
    \bottomrule
    \end{tabular}
\end{table}
\paragraph{Training Method. }
In addition to evidence-conditioned self-distillation (ECSD), we explored supervised fine-tuning (SFT)~\cite{SFT} and two subsequent policy-optimization objectives.  As illustrated in Table~\ref{tab:train}, direct SFT on golden answers performs substantially worse than ECSD. 
Starting from the SFT checkpoint, both on-policy distillation (OPD)~\cite{li2026rethinkingonpolicydistillationlarge} and group relative policy optimization (GRPO)~\cite{shao2024deepseekmathpushinglimitsmathematical} improve cross-domain LoCoMo performance, reaching 52.80 and 53.18 F1, respectively. However, neither matches ECSD, which achieves 54.11 F1.
The best GRPO variant reaches 53.18 $F_1$, compared with 54.11 for ECSD.  This result suggests that answer-level reinforcement learning does not compensate for a weaker memory-representation objective: ECSD's token-level supervision from a privileged-evidence teacher remains important for learning the TransMem module.

\section{Conclusion}
We propose TransMem, a lightweight parametric memory module that improves test-time long-context reasoning without explicit memory construction or backbone scaling. TransMem transforms sparse, position-sensitive historical hidden states into reusable memory representations and injects them into current reasoning. Evidence-conditioned self-distillation trains the module to recover an evidence-only teacher’s predictions from full contexts. Experiments on LoCoMo, MemoryAgentBench, and HotpotQA show consistent gains across backbones with sparse memory representations and context-independent overhead, supporting efficient decoupling of memory capability from backbone reasoning capacity. Future work will explore adaptive MoE memory extensions for TransMem and applications to long-horizon agent scenarios.

%% file: tab/main.tex
\providecommand{\scoredelta}[2]{#1\textsubscript{\textcolor{black!90}{#2}}}
\definecolor{mindlabbg}{RGB}{245,245,245}
\begin{table*}[t]
    \centering
    \caption{\textbf{Main benchmark results comparing different memory mechanisms.}}
    \resizebox{\textwidth}{!}{
    \begin{tabular}{lccccccc}
    \toprule
    \multirow{2}{*}{\textbf{Model}} 
    & \multicolumn{5}{c}{\textbf{LoCoMo}} 
    & \multicolumn{2}{c}{\textbf{HotpotQA}} \\
    \cmidrule(lr){2-6}
    \cmidrule(lr){7-8}
    & \textbf{Avg.} 
    & \textbf{Multi-Hop} 
    & \textbf{Temporal} 
    & \textbf{OpenDomain} 
    & \textbf{Single-Hop}
    & \textbf{EM}
    & \textbf{F1} \\
    \midrule
    \textbf{Qwen3-4B-Instruct} & 40.79 & 38.39 & 32.89 & 10.77 & 48.05 & 42.35 & 56.00 \\
    \rowcolor{mindlabbg} \multicolumn{8}{c}{\textit{External Memory}} \\
    \quad + BM25 RAG & \scoredelta{36.68}{-4.11} & \scoredelta{38.12}{-0.27} & \scoredelta{20.34}{-12.55} & \scoredelta{9.99}{-0.78} & \scoredelta{45.47}{-2.58} & \scoredelta{40.35}{-2.00} & \scoredelta{52.83}{-3.17} \\
    \quad + LLMLingua-2 & \scoredelta{40.98}{+0.19} & \scoredelta{39.07}{+0.68} & \scoredelta{30.13}{-2.76} & \scoredelta{10.98}{+0.21} & \scoredelta{49.19}{+1.14} & \scoredelta{36.93}{-5.42} & \scoredelta{50.03}{-5.97} \\
    \quad + MemoryBank & \scoredelta{38.14}{-2.65} & \scoredelta{37.88}{-0.51} & \scoredelta{21.76}{-11.13} & \scoredelta{13.35}{+2.58} & \scoredelta{47.31}{-0.74} & \scoredelta{39.16}{-3.19} & \scoredelta{51.25}{-4.75}\\
    \rowcolor{mindlabbg} \multicolumn{8}{c}{\textit{Parametric Memory}} \\
    \quad + Context2LoRA & \scoredelta{48.11}{+7.32} & \scoredelta{37.95}{-0.44} & \scoredelta{34.99}{+2.10} & \scoredelta{16.75}{+5.98} & \scoredelta{60.11}{+12.06} & \scoredelta{37.85}{-4.50} & \scoredelta{50.88}{-5.12} \\
    \quad + MemGen & \scoredelta{40.05}{-0.74} & \scoredelta{32.93}{-5.46} & \scoredelta{33.30}{+0.41} & \scoredelta{12.67}{+1.90} & \scoredelta{48.13}{+0.08} & \scoredelta{5.36}{-36.99} & \scoredelta{16.27}{-39.73} \\
    \quad + MLP Memory & \scoredelta{40.21}{-0.58} & \scoredelta{32.87}{-5.52} & \scoredelta{21.72}{-11.17} & \scoredelta{13.81}{+3.04} & \scoredelta{52.75}{+4.70} & \scoredelta{44.92}{+2.57} & \scoredelta{59.23}{+3.23} \\
    \quad + $\delta$-Mem & \scoredelta{46.53}{+5.74} & \scoredelta{42.14}{+3.75} & \scoredelta{37.20}{+4.31} & \scoredelta{13.35}{+2.58} & \scoredelta{55.36}{+7.31} & \scoredelta{\textbf{49.41}}{+7.06} & \scoredelta{63.66}{+7.66}\\
    \midrule
    \quad + \textbf{TransMem (Ours)} & \scoredelta{\textbf{54.11}}{+13.32} & \scoredelta{\textbf{52.04}}{+13.65} & \scoredelta{\textbf{45.62}}{+12.73} & \scoredelta{\textbf{20.64}}{+9.87} & \scoredelta{\textbf{61.87}}{+13.82} & \scoredelta{48.55}{+6.20} & \scoredelta{\textbf{66.20}}{+10.20}\\
    \midrule
    \textbf{Qwen2.5-14B-Instruct} & 42.74 & 32.59 & 27.36 & 13.50 & 55.35 & 50.75& 60.20\\
    \rowcolor{mindlabbg} \multicolumn{8}{c}{\textit{External Memory}} \\
    \quad + BM25 RAG & \scoredelta{42.25}{-0.49} & \scoredelta{32.17}{-0.42} & \scoredelta{30.77}{+3.41} & \scoredelta{23.21}{+9.71} & \scoredelta{52.19}{-3.16} & \scoredelta{50.46}{-0.29} & \scoredelta{61.2}{+1.00}\\
    \quad + LLMLingua-2 & \scoredelta{43.43}{+0.69} & \scoredelta{34.12}{+1.53} & \scoredelta{31.15}{+3.79} & \scoredelta{19.48}{+5.98} & \scoredelta{53.98}{-1.37} & \scoredelta{45.45}{-5.30} & \scoredelta{59.10}{-1.10}\\
    \quad + MemoryBank & \scoredelta{44.64}{+1.90} & \scoredelta{34.50}{+1.91} & \scoredelta{\textbf{44.70}}{+17.34} & \scoredelta{23.86}{+10.36} & \scoredelta{50.39}{-4.96} & \scoredelta{46.66}{-4.09} & \scoredelta{58.51}{-1.69}\\
    \rowcolor{mindlabbg} \multicolumn{8}{c}{\textit{Parametric Memory}} \\
    \quad + Context2LoRA & \scoredelta{48.28}{+5.54} & \scoredelta{35.16}{+2.57} & \scoredelta{28.34}{+0.98} & \scoredelta{17.23}{+3.73} & \scoredelta{63.85}{+8.50} & \scoredelta{48.15}{-2.60} & \scoredelta{59.98}{-0.22}\\
    \quad + MemGen & \scoredelta{43.64}{+0.90} & \scoredelta{29.36}{-3.23} & \scoredelta{29.47}{+2.11} & \scoredelta{17.54}{+4.04} & \scoredelta{56.83}{+1.48} & \scoredelta{10.56}{-40.19} & \scoredelta{21.45}{-38.75}\\
    \quad + MLP Memory & \scoredelta{45.33}{+2.59} & \scoredelta{35.65}{+3.06} & \scoredelta{25.46}{-1.90} & \scoredelta{21.39}{+7.89} & \scoredelta{58.90}{+3.55} & \scoredelta{47.35}{-3.40} & \scoredelta{58.84}{-1.36}\\
    \quad + $\delta$-Mem & \scoredelta{47.83}{+5.09} & \scoredelta{34.99}{+2.40} & \scoredelta{27.47}{+0.11} & \scoredelta{27.37}{+13.87} & \scoredelta{62.25}{+6.90} & \scoredelta{51.38}{+0.63} & \scoredelta{65.36}{+5.16}\\
    \midrule
    \quad + \textbf{TransMem (Ours)} & \scoredelta{\textbf{54.32}}{+11.58} & \scoredelta{\textbf{41.20}}{+8.61} & \scoredelta{41.44}{+14.08} & \scoredelta{\textbf{30.38}}{+16.88} & \scoredelta{\textbf{66.38}}{+11.03} & \scoredelta{\textbf{58.01}}{+7.26} & \scoredelta{\textbf{72.34}}{+12.14}\\
    \midrule
    \textbf{Llama3.1-8B-Instruct} & 22.39& 20.30 & 19.42 & 25.86 & 23.84 & 49.77& 58.60\\
    \rowcolor{mindlabbg} \multicolumn{8}{c}{\textit{External Memory}} \\
    \quad + BM25 RAG & \scoredelta{19.25}{-3.14} & \scoredelta{12.28}{-8.02} & \scoredelta{11.58}{-7.84} & \scoredelta{31.02}{+5.16} & \scoredelta{23.18}{-0.66} & \scoredelta{44.95}{-4.82} & \scoredelta{55.12}{-3.48}\\
    \quad + MemoryBank & \scoredelta{23.53}{+1.14} & \scoredelta{23.31}{+3.01} & \scoredelta{13.01}{-6.41} & \scoredelta{31.05}{+5.19} & \scoredelta{26.77}{+2.93} & \scoredelta{50.15}{+0.38} & \scoredelta{65.97}{+7.37}\\
    \quad + SimpleMem & \scoredelta{27.47}{+5.08} & \scoredelta{30.44}{+10.14} & \scoredelta{18.77}{-0.65} & \scoredelta{31.23}{+5.37} & \scoredelta{29.37}{+5.53} & \scoredelta{52.06}{+2.29} & \scoredelta{66.15}{+7.55}\\
    \quad + Memory-R1-GRPO & \scoredelta{26.84}{+4.45} & \scoredelta{32.87}{+12.57} & \scoredelta{16.72}{-2.70} & \scoredelta{8.81}{-17.05} & \scoredelta{30.75}{+6.91} & \scoredelta{48.54}{-1.23} & \scoredelta{62.15}{+3.55}\\
    \quad + TriMem & \scoredelta{38.70}{+16.31} & \scoredelta{34.56}{+14.26} & \scoredelta{32.36}{+12.94} & \scoredelta{\textbf{32.71}}{+6.85} & \scoredelta{43.20}{+19.36} & \scoredelta{55.36}{+5.59} & \scoredelta{68.11}{+9.51}\\
    \rowcolor{mindlabbg} \multicolumn{8}{c}{\textit{Parametric Memory}} \\
    \quad + $\delta$-Mem & \scoredelta{46.21}{+23.82} & \scoredelta{36.64}{+16.34} & \scoredelta{31.28}{+11.86} & \scoredelta{24.87}{-0.99} & \scoredelta{57.56}{+33.72} & \scoredelta{56.27}{+6.50} & \scoredelta{69.47}{+10.87}\\
    \midrule
    \quad + \textbf{TransMem (Ours)} & \scoredelta{\textbf{51.64}}{+29.25} & \scoredelta{\textbf{47.27}}{+26.97} & \scoredelta{\textbf{37.29}}{+17.87} & \scoredelta{28.69}{+2.83} & \scoredelta{\textbf{61.21}}{+37.37} & \scoredelta{\textbf{56.57}}{+6.80} & \scoredelta{\textbf{71.63}}{+13.03}\\
    \bottomrule
    \end{tabular}
    }
    \label{tab:main}
\end{table*}


%% file: tab/MAB.tex
\definecolor{mindlabbg}{RGB}{245,245,245}
\begin{table}[t]
    \centering
    \caption{Results on Memory Agent Bench comparing different memory mechanisms.}
    \resizebox{0.48\textwidth}{!}{
    \begin{tabular}{lccccc}
    \toprule
    \textbf{Model} & \textbf{Avg.} & \textbf{AR} & \textbf{TTL} & \textbf{LRU} & \textbf{SF} \\
    \midrule
    \textbf{Qwen3-4B-Instruct} & 29.54
& 35.30 & 26.14 & \textbf{47.08} & 14.37 \\
    \rowcolor{mindlabbg} \multicolumn{6}{c}{\textit{External Memory}} \\
    \quad + BM25 RAG & 24.43
& 32.20 & 9.74 & 37.86 & 15.00 \\
    \quad + LLMLingua-2 & 15.63
& 21.45 & 1.43 & 38.45 & 8.62 \\
    \quad + MemoryBank  & 17.65
& 22.65 & 7.67 & 36.36 & 9.88 \\
\rowcolor{mindlabbg} \multicolumn{6}{c}{\textit{Parametric Memory}} \\
    \quad + Context2LoRA & 32.53
& 40.00 & 29.86 & 25.15 & 17.75 \\
    \quad + MemGen   & 29.61
& 34.85 & 28.45 & 44.30 & 14.38 \\
    \quad + MLP Memory  & 28.80
& 35.35 & 26.00 & 31.19 & 14.38 \\
    \quad + $\delta$-Mem & 38.85
& \textbf{44.40} & 47.29 & 41.55 & 17.00 \\
    \midrule
    \quad + \textbf{TransMem (Ours)}  & \textbf{40.00}& 41.60& \textbf{54.64}& 36.19& \textbf{24.00}\\
    \bottomrule
    \end{tabular}
    }
    \label{tab:MAB}
\end{table}



%% file: supp.tex
\section{More Experiments}

\subsection{Effect of the Train Set}
We next compare different training corpora while fixing the student to
Qwen3-4B and inserting dynamically gated TransMem modules into its last
four layers.  We evaluate all checkpoints on the same 1,540 questions
from LoCoMo and report token-level $F_1$ in Table
\ref{tab:training-corpora}.

The HotpotQA--LongMemEval mixture reaches 53.95 $F_1$, only $0.16$
points below the main HotpotQA recipe.  In contrast, aggressively
downsampling HotpotQA and adding the small LoCoMo-train split
reduces overall $F_1$ to 51.80, with the largest degradation on temporal
questions.  The last train set contains 84 LoCoMo questions that
also occur in the full evaluation set; removing them at evaluation
gives 51.87 $F_1$, essentially the same conclusion.  Hence, the decrease
is not explained by evaluation overlap.  Since the last recipe changes
both corpus composition and the number of HotpotQA examples, the
results do not show that LoCoMo supervision is intrinsically harmful;
rather, this particular mixture provides no improvement.

\begin{table}[htbp]
  \centering
  \normalsize
  \setlength{\tabcolsep}{2.5pt}
  \caption{LoCoMo $F_1$ with different training corpora for
  Qwen3-4B TransMem.  HQA, LME, and LoCoMo-T denote HotpotQA,
  LongMemEval, and LoCoMo-train.  MH, Temp., Open, and SH denote
  multi-hop, temporal, open-domain, and single-hop questions.}
  \begin{tabular}{@{}lccccc@{}}
    \toprule
    Train set & Overall & MH & Temp. & Open & SH \\
    \midrule
    HQA
      & \textbf{54.11} & \textbf{52.04} & \textbf{45.62} & 20.64 & 61.87\\
    HQA + LME
      & 53.95& 50.40 & 45.54& \textbf{21.85} &
        \textbf{62.02} \\
    \shortstack[l]{15\% HQA + LME\\+ LoCoMo-T}
      & 51.80 & 50.53 & 37.20 & 19.13 & 61.53 \\
    \bottomrule
  \end{tabular}
  \label{tab:training-corpora}
\end{table}

\subsection{Gate Parameterization}
We compare three gate parameterizations on Qwen3-4B with
the Last-4 configuration, keeping HotpotQA as the train set.  Let
$z=W_g h^m/\tau$.  Besides a fixed residual scale, we evaluate the
non-negative centered sigmoid used by our main model,
$g=2\sigma(z)$, and a signed alternative,
$g=3\sigma(z)-1$.  The latter can suppress the residual with a
negative correction.

\begin{table}[htbp]
  \centering
  \caption{LoCoMo $F_1$ under different gate parameterizations.
  All dynamic gates are initialized at $g=1$.}
  \begin{tabular}{lccccc}
    \toprule
    Gate & Overall & MH & Temp. & Open & SH \\
    \midrule
    $g=1$
      & 53.09& 48.04& 45.09& \textbf{21.57}& 61.45\\
    $g=2\sigma(z)$
      & \textbf{54.11} & \textbf{52.04} & 45.62 & 20.64 &
        \textbf{61.87} \\
    $g=3\sigma(z)-1$
      & 54.01 & 51.47 & \textbf{45.87} & 20.87 & 61.76 \\
    \bottomrule
  \end{tabular}
  \label{tab:gate-parameterization}
\end{table}

Both learned gates improve over the fixed residual scale.  The main
$2\sigma(z)$ gate performs best overall, improving the fixed-gate model
by $0.46$ $F_1$ points.  Allowing negative corrections does not provide an
additional benefit: $3\sigma(z)-1$ is $0.10$ points below the main
parameterization, although it attains the best temporal score.  We
therefore use $2\sigma(z)$ for its slightly stronger overall result and
its simpler non-negative modulation.

\subsection{Early Final-Hidden TransMem}
Before introducing layer-wise injection, we studied a final-hidden
variant that applies a single TransMem module after the backbone's
final normalization and before the language-model head.  Unlike the
current method, the module itself contains either two or four
Transformer blocks, denoted by $L_{\mathrm{TM}}$.  It uses a fixed
residual scale and has no learned gate.  Table
\ref{tab:qwen3-4b-early-final-hidden} summarizes this early architecture on
held-out HotpotQA examples.

\begin{table}[t]
  \centering
  \caption{HotpotQA results (\%) with Qwen3-4B-Instruct
  (the 2507 checkpoint).
  The first five rows use Final-hidden and report Exact
  Match/Contains (E/C) on the archived 128-example split;
  $L_{\mathrm{TM}}$ is the number of Transformer blocks inside its
  single memory module.  The final row gives the main TransMem result
  under the paper's official EM/$F_1$ protocol and is included as a
  reference rather than a directly controlled comparison.}
  \resizebox{0.48\textwidth}{!}{
  \begin{tabular}{cllc}
    \toprule
    Method & Memory layout & Train set & HQA score \\
    \midrule
    Final-hidden & $L_{\mathrm{TM}}=4$ & HQA+LME & 39.8/50.8 \\
    Final-hidden & $L_{\mathrm{TM}}=2$ & HQA     & 39.1/50.8 \\
    Final-hidden & $L_{\mathrm{TM}}=2$ & HQA+LME & 39.1/48.4 \\
    Final-hidden & $L_{\mathrm{TM}}=4$ & LME     & 36.7/50.0 \\
    Final-hidden & $L_{\mathrm{TM}}=2$ & LME     & 38.3/52.3 \\
    \midrule
    \textbf{TransMem (Ours)} & Last-4 & HQA & \textbf{48.55/66.20} \\
    \bottomrule
  \end{tabular}
  }
  \label{tab:qwen3-4b-early-final-hidden}
\end{table}

Changing the internal depth from two to four does not yield a
consistent improvement on HotpotQA.  The two mixed-corpus
configurations with $L_{\mathrm{TM}}=4$ and $L_{\mathrm{TM}}=2$ also
differ in training duration, so they should not be interpreted as a
strictly controlled depth comparison.  More broadly, these results
suggest that merely scaling up the memory parameters at a single
final-hidden location has limited effect.  This observation motivated
the current design: instead of concentrating capacity at one location,
we inject small memory modules at multiple LLM layers so that memory
corrections can interact with different stages of the backbone
computation.

\subsection{Additional Hidden-State Supervision}
We also tested whether directly matching the teacher's final-layer
feature improves knowledge transfer.  In addition to the
self-distillation objective, the Last-4 model was trained with
\[
  \mathcal{L}
  =
  \mathcal{L}_{\mathrm{SD}}
  + \beta
  \left\lVert
    H'_{\mathrm{last}}-H_{\mathrm{last}}^{\mathrm{tea}}
  \right\rVert_2^2 .
\]
Here, $H'_{\mathrm{last}}$ is the answer-position feature from the
student LLM's final layer after the TransMem update, and
$H_{\mathrm{last}}^{\mathrm{tea}}$ is the corresponding final-layer
feature produced by the privileged-evidence teacher.  The prime
therefore denotes the TransMem-updated student representation.
All variants use Qwen3-4B and the same QASPER--HotpotQA
multi-domain train set.

\begin{table}[t]
  \centering
  \caption{LoCoMo $F_1$ when adding hidden-state regression to
  the self-distillation objective. All rows use the Qwen3-4B Last-4
  architecture.}
  \begin{tabular}{cccccc}
    \toprule
    $\beta$ & Overall & MH & Temp. & Open & SH \\
    \midrule
    0   & \textbf{54.11} & \textbf{52.04} & 45.62 & 20.64 & \textbf{61.87} \\
    0.3 & 51.91 & 46.94& 44.12 & 19.16 & 60.28 \\
    1.0 & 52.18& 46.38 & 44.26& 19.29 & 60.90 \\
    3.0 & 51.79 & 45.29 & 43.62 & \textbf{20.89} & 60.62 \\
    \bottomrule
  \end{tabular}
  \label{tab:hidden-state-supervision}
\end{table}

Hidden-state supervision provides no benefit in this ablation: every
$\beta>0$ setting performs below the no-regression variant
($\beta=0$).  One likely explanation is that the output-distribution
objective already supplies the task-relevant behavioral signal,
whereas imposing additional Euclidean alignment over-constrains the
student's hidden space.  In other words, supervising hidden states too
strongly may restrict functionally equivalent representations without
improving cross-domain memory use.

\subsection{TransMem Feature Source}
Our default Last-4 model computes the memory update for LLM layer $K$
from the output $H^K$ of that layer.  We compare it with
\emph{transmem-before}, which computes the same update from the layer
input $H^{K-1}$ while still adding the resulting residual to $H^K$.
All other architectural and training choices remain unchanged.

\begin{table}[t]
  \centering
  \caption{Feature-source ablation for Qwen3-4B Last-4 TransMem.}
  \begin{tabular}{lccc}
    \toprule
    Variant & Source state & Added to & LoCoMo $F_1$ \\
    \midrule
    Default & $H^K$ & $H^K$ & \textbf{54.11} \\
    Before & $H^{K-1}$ & $H^K$ & 53.92 \\
    \bottomrule
  \end{tabular}
  \label{tab:transmem-before}
\end{table}

Reading the post-block state is 0.19 $F_1$ points better, although the
small gap indicates that TransMem is not highly sensitive to this
choice.  We retain $H^K$ because it exposes each memory module to the
representation already transformed by its corresponding LLM layer.

\subsection{Gate Statistics on Qwen3-4B}
Table \ref{tab:gate-statistics} compares the learned gate distributions
under architectural, parameterization, and train-set changes.  Every
experiment in this table uses Qwen3-4B.  The reported mean and standard
deviation aggregate validation answer positions across all injected
layers; because the gate is token-dependent, they are descriptive
statistics rather than fixed model parameters.

\begin{table}[t]
  \centering
  \caption{Gate statistics and LoCoMo $F_1$ for Qwen3-4B.
  HQA, LME, and LoCoMo-F denote HotpotQA, LongMemEval, and the
  full LoCoMo train set.}
  \begin{tabular}{lllcc}
    \toprule
    Setting & Train set & Gate & $\mu/\sigma$ & $F_1$ \\
    \midrule
    Last-4 & HQA & $2\sigma(z)$ & 1.137/0.384 & \textbf{54.11} \\
    Last-8 & HQA & $2\sigma(z)$ & 0.605/0.319 & 53.27 \\
    Last-4 & HQA & $3\sigma(z)-1$ & 1.200/0.398 & 54.01 \\
    \shortstack[l]{Last-4\\(before)}
      & HQA & $2\sigma(z)$ & 1.237/0.313 & 53.92 \\
    Last-4 & \shortstack[l]{LME+\\LoCoMo-F}
      & $2\sigma(z)$ & 1.561/0.143 & 50.68 \\
    \bottomrule
  \end{tabular}
  \label{tab:gate-statistics}
\end{table}

The Last-4 and Last-8 models learn qualitatively different global
scales: Last-4 mildly amplifies memory updates on average, whereas
Last-8 suppresses them.  The signed gate and transmem-before variants
remain close to the main model.  Training on the much smaller
LME--LoCoMo mixture produces a high-mean, low-variance gate and a lower
$F_1$ despite containing target-domain examples, consistent with
overfitting rather than improved memory use.

\subsection{Layer-Wise Gate Behavior}
\begin{figure}[htbp]
  \centering
  \includegraphics[width=\columnwidth]{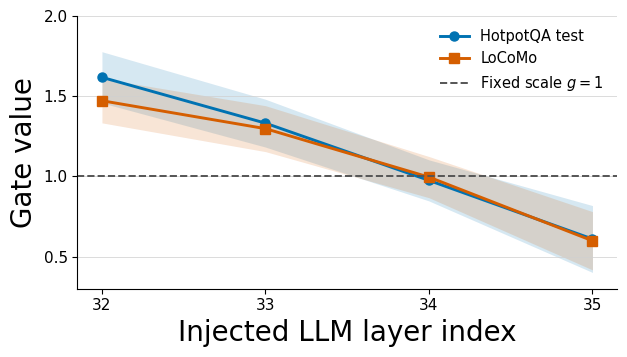}%
  \caption{Layer-wise gate values for Qwen3-4B Last-4 TransMem.
  Curves show means and shaded regions show one standard deviation
  across answer tokens. The dashed line is the fixed residual scale
  $g=1$.}
  \label{fig:gate-layerwise}
\end{figure}

Aggregate statistics hide a consistent division of labor among the
four injected layers.  Figure \ref{fig:gate-layerwise} reports the
token-level gate distribution for the main Qwen3-4B Last-4 checkpoint
on 128 in-domain HotpotQA development questions and a
386-question cross-domain LoCoMo diagnostic subset.

The gate forms a monotonic amplification-to-suppression profile:
Layer 32 strongly amplifies the TransMem update, Layers 33--34
progressively reduce it, and Layer 35 suppresses the direct
intervention.  More importantly, the same profile is preserved on
LoCoMo.  The corresponding means from Layers 32 to 35 are
$1.618,1.332,0.975,0.610$ on HotpotQA and
$1.472,1.298,0.996,0.599$ on LoCoMo.  This cross-domain stability
suggests that the gate learns a layer-specific correction schedule
rather than a dataset-specific binary switch.

\subsection{More Models}
We additionally train Last-4 TransMem with HotpotQA on two other
student backbones.  Table \ref{tab:more-models} reports their absolute
LoCoMo scores together with Qwen3-4B as a reference.

\begin{table}[t]
  \centering
  \caption{LoCoMo $F_1$ with additional student backbones.}
  \begin{tabular}{lccc}
    \toprule
    Student backbone & Injection & Train set & $F_1$ \\
    \midrule
    Qwen2.5-7B & Last-4 & HQA & 49.98 \\
    Qwen3-8B & Last-4 & HQA & 50.76 \\
    \bottomrule
  \end{tabular}
  \label{tab:more-models}
\end{table}

The method can be trained without changing the Last-4 design across
these model sizes and families.  Absolute performance does not scale
monotonically with parameter count under this protocol, so these
results support architectural portability rather than a claim that a
larger frozen student necessarily yields a larger gain.

\subsection{Sensitivity to the Number of Memory Segments}
We first study the memory-segmentation granularity, denoted by $B$ in the paper.  We reuse the Qwen3-4B TransMem checkpoint trained with $B=4$ and vary only $B$ at inference time.  Thus, this experiment
measures inference-time sensitivity rather than the effect of retraining a separate model for every value of $B$.  Table \ref{tab:memory-segments} reports the official task metrics (percentage points) on the deterministic subsets of MemoryAgentBench.

The results are nearly invariant over $B\in\{4,8,16,32\}$: increasing
$B$ from 4 to 32 changes EventQA by only $0.2$ points and improves
RULER-1 by $1.0$ point, while the other three scores remain unchanged.
This suggests that TransMem is not brittle to the precise segmentation
granularity.  We therefore retain $B=4$ as the default because it uses
fewer memory states.

\begin{table}[htbp]
  \centering
  \caption{Inference-time sensitivity to the number of memory
  segments $B$.  FC-SH and FC-MH denote the single-hop and multi-hop
  FactConsolidation subsets, respectively.}
  \begin{tabular}{c@{\hspace{7pt}}ccccc}
    \toprule
    $B$ & RULER-1 & RULER-2 & EventQA & FC-SH & FC-MH \\
    \midrule
    4  & 40.00 & \textbf{30.00} & 47.40 & \textbf{43.00} & \textbf{5.00} \\
    8  & 40.00 & \textbf{30.00} & 47.20 & \textbf{43.00} & \textbf{5.00} \\
    16 & 40.00 & \textbf{30.00} & 47.20 & \textbf{43.00} & \textbf{5.00} \\
    32 & \textbf{41.00} & \textbf{30.00} & \textbf{47.60} &
         \textbf{43.00} & \textbf{5.00} \\
    \bottomrule
  \end{tabular}
  \label{tab:memory-segments}
\end{table}